\documentstyle[aps,preprint]{revtex}
\begin{document}
\draft
\preprint{\vbox
{\hbox{SNUTP 95-050} \hbox{hep-ph/9505400}}}
\title{Normalization of the Perturbative QCD Corrections\\
       for $B\rightarrow X_s \, \gamma$ Decay}

\author{Junegone Chay\footnote{E-mail address:
\tt chay@kupt.korea.ac.kr}
and Soo-Jong Rey\footnote{E-mail address:
\tt sjrey@phyb.snu.ac.kr}}

\address{Physics Department, Korea University, Seoul 136-701,
Korea${}^*$ \\
  Physics Department and Center for Theoretical Physics \\
  Seoul National University, Seoul 151-752, Korea${}^\dagger$ }

\maketitle

\begin{abstract}
We study the normalization of perturbative QCD corrections to the
inclusive $B \rightarrow X_s \gamma$ decay. We propose to set the 
renormalization scale using the Brodsky-Lepage-Mackenzie (BLM) method. 
In the proposed method the scale is determined by absorbing the vacuum 
polarization correction from light fermions to renormalization scale 
but not including the anomalous dimensions. 
The BLM scale depends in general on the renormalization scheme and the 
factorization scale. We find that the BLM scale is insensitive to the
factorization scale. In the heavy-quark potential scheme, we find that 
the BLM scale is $\mu_{BLM} \approx (0.315 - 0.334)  m_b$ when the
factorization scale varies from $m_b/2$ to $2 m_b$. 
\end{abstract}
The rare decay process $b \rightarrow s 
\gamma$ has been recently observed at CLEO II\cite{cleo2} at a
branching ratio $ 2.32 (\pm 0.57 \pm 0.35) \times 10^{-4}$
(statistical and systematic errors). The process serves as an
excellent probe to the new physics beyond the Standard Model such as
charged Higgs\cite{chargedhiggs}, supersymmetry\cite{susy} or anomalous
$WW\gamma$ coupling\cite{wwgamma}. At leading order the process is
described by an electromagnetic penguin diagram. It is
known that there are large QCD corrections to the penguin diagram. A
complete renormalization group-improved perturbative calculation is
now available to the leading-logarithm approximation. It is found that
the QCD corrections increase the rate by a factor two to three. 
In the Standard Model the branching ratio to leading-logarithmic 
accuracy is $2.8 (\pm 0.8)
\times 10^{-4}$, where the error is dominated by the uncertainty in
the QCD renormalization scale varied over $m_b/2 < \mu < 2 m_b$
\cite{buras,martinelli}. Inclusion of the next-to-leading logarithmic 
QCD corrections will reduce the theoretical error. However, a complete
calculation is still not achieved.  With partial next-to-leading 
corrections known so far, the branching ratio falls to $\sim 1.9 
\times 10^{-4}$.  

In view of the current status of the theoretical uncertainties as
summarized above it is of interest to normalize the leading order
QCD corrections and estimate theoretical errors thereof. In this
paper we address this issue and determine the renormalization scale
according to the Brodsky-Lepage-Mackenzie (BLM) method \cite{blm}. So
far applications of the BLM method have been mostly restricted to 
processes involving (partially) conserved currents. In this case 
QCD correction may be calculated by a perturbative expansion at 
a fixed order. For example, BLM scale setting has been
established for the inclusive semileptonic $B$ decay $B \rightarrow
X_q e {\overline \nu}$ \cite{lukesavagewise}. In 
Ref.~\cite{lukesavagewise} the BLM scale in the $\overline {MS}$ scheme
is found $\mu_{BLM} \approx 0.07 \, m_b$ ($0.12 \, m_b$ when running
$\overline{MS}$ b-quark mass \cite{broadmass} is used), indicating a
significant QCD correction from higher order terms. The QCD
corrections for the radiative $B$ decay is more complicated since the
process involves large logarithms arising from QCD effects between the
top-quark or $W$-gauge boson mass and the bottom quark mass
scales. 
The effective Hamiltonian approach takes care of these large logarithms
by introducing a factorization scale and separates the long- and the 
short-distance QCD effects.

In situations involving large logarithms a physical observable $P$ 
such as the total decay rate or the branching ratio may be written
schematically as 
\begin{eqnarray}
P &=& M(\mu) \Big[ r_0 (\mu) + r_1 (\mu, \mu_R) 
{\alpha_s (\mu_R) \over \pi}  \nonumber \\
&+& n_f r_2 (\mu, \mu_R) \Big({\alpha_s (\mu_R) \over \pi} \Big)^2 
+ \cdots \Big].
\label{schem}
\end{eqnarray}
Here $\mu$ is a factorization scale dividing long- and short-distance 
physics and $\mu_R$ is a renormalization scale for
the QCD corrections. All the large logarithms of $(m_W/\mu)$ are 
resummed into $M(\mu)$. QCD corrections to the matrix element at 
$\mu, \mu_R \gg \Lambda_{QCD}$ are given by a perturbative expansion. 
The coefficients $r_0, r_1$ and $r_2$ depend on both the factorization 
and the renormalization scales. 
Of course the physical observable $P$ calculated to all orders should 
be independent of the two scales. 
However, at any finite order, perturbative evaluation of $P$
do depend on the factorization and the renormalization scales. 
In practice the factorization scale is set equal to the renormalization
scale, and then the theoretical error is estimated. To be precise
these two scales are distinct, and there is no reason why
the two scales should be set equal. For example, in the conformally
invariant QCD, a finite-order perturbatiion expansion does not
depend on any renormalization scale but still depends explicitly 
on the factorization scale.

We are interested in calculating perturbative QCD corrections and
setting the renormalization scale $\mu_R$ for a given renormalization
scheme and the factorization scale $\mu$. Typically there are
logarithms of $(\mu_R / \mu)$ present from two sources. One is the
logarithms that can be resummed to the running coupling
constants. Another is the logarithms that may be resummed into the
anomalous dimensions of operators involved. In order to set the
renormalization scale we propose to follow the original spirit of BLM
closely. The idea is to set $\mu_R$ by absorbing $n_f \alpha_s^2$
contributions of the vacuum polarization to the renormalization scale,
but not including the contributions of the anomalous dimensions. 
Then the generalized BLM scale is set by the fixed-order perturbative
correction part inside the bracket in Eq.(1): 
$\mu_{BLM} = \mu_R \, \exp (3 r_2 / r_1)$. 
If the content of the theory does not change between $\mu$ and $\mu_R$,
the normalized physical quantity ${\overline P}$ may be given by
\begin{equation}
{\overline P} = M(\mu_{\rm BLM})
\Big[ r_0 (\mu_{BLM}) + r_1( \mu_{BLM}) {\alpha_s (\mu_{\rm BLM})
\over \pi} \Big] 
\label{norm}
\end{equation}
in which the anomalous dimension effects are taken into account by
running from $M(\mu)$ to $M(\mu_{BLM})$ using the renormalization group 
equation. Aside from setting the BLM scale it may be also argued that the
fixed order coefficients $r_1, r_2$ are the most significant parts in
the next-to-leading order contribution \cite{martinelli,aligreub}.

We emphasize, however, that our proposal is by no means unique or
superior to other plausible generalizations of the BLM method. 
For example, we may set the BLM scale including contributions 
from the next-to-leading order anomalous dimensions as well as effective 
coupling constant through the vacuum polarization. This is complicated
and in order to be consistent, we need anomalous dimensions at
next-to-leading order. Furthermore at present a complete
next-to-leading order calculation of the coefficient functions and
anomalous dimensions is not available for the radiative $B$ decay.
On the other hand our method provides a physically
motivated generalization of the original BLM method and the 
final result shows a reasonable behavior.

For the radiative $B$ decay, the effective Hamiltonian below $M_W$
scale is given by \cite{effhamil}
\begin{equation}
H_{\mbox{eff}} = -V_{tb}V^*_{ts} \, \frac{G_F}{\sqrt{2}} \sum_{i=1}^8 
C_i(\mu) \, {\cal O}_i (\mu),
\end{equation}
where $\mu$ is the factorization scale separating the coefficient
functions and the operators. For the partonic decays $b\rightarrow
s\gamma$ and $b\rightarrow s\gamma g$, the dominant contributions in
the radiative decay arise from the
operators \cite{effhamil}
\begin{eqnarray}
{\cal O}_1 &=& ({\overline c}_{L\alpha} \gamma^\mu b_{L \beta}) 
({\overline s}_{L \beta} \gamma_\mu c_{L \alpha}) \nonumber \\
{\cal O}_2 &=& ({\overline c}_{L\alpha} \gamma^\mu b_{L \alpha}) 
({\overline s}_{L \beta} \gamma_\mu c_{L \beta}) \nonumber \\
{\cal O}_7 &=& {e \over 16 \pi^2} F_{\mu \nu}  {\overline s}_\alpha  
\sigma^{\mu \nu} (m_b P_R + m_s P_L) b_\alpha \nonumber \\
{\cal O}_8 &=& {g_s \over 16 \pi^2} G_{\mu \nu}^A  {\overline
s}_\alpha \sigma^{\mu \nu} T_{\alpha \beta}^A (m_b P_R + m_s P_L)
b_\beta 
\label{ops}
\end{eqnarray}
where $\alpha, \beta$ are color indices, $P_{R,L} = (1 \pm
\gamma_5)/2$ and $F_{\mu \nu}, G_{\mu \nu}^A$ are photon and gluon
field strengths respectively. Contributions of the other operators
enter through operator mixing at next-to-leading order and are
numerically negligible\cite{aligreub}. Because of color structure,
${\cal O}_1$ does not contribute to the real gluon emission to the
order we calculate. In addition $C_8 (m_b)$ is small compared to the
others. Hence we neglect ${\cal O}_1, \, {\cal O}_8$ contributions in
what follows.  

The Wilson coefficients at leading order are given by\cite{effhamil}
\begin{eqnarray}
C_2 (\mu)&=& \frac{1}{2}  \Bigl[  \eta^{-6/23} + \eta^{12/23}  \Bigr] 
 C_2(m_W), 
\nonumber \\
C_7 (\mu) &=& \eta^{-16/23}  \Bigl\{ C_7(m_W)
-\frac{58}{135}  \Bigl[  \eta^{10/23} -1  \Bigr]  C_2(m_W) \nonumber
\\ 
&-&\frac{29}{189}  \Bigl[  \eta^{28/23} -1  \Bigr]  C_2(m_W)\Bigr\},
\label{coeff}
\end{eqnarray}
with $\eta = \alpha_s (\mu)/\alpha_s(m_W)$ and 
\begin{eqnarray}
C_2 (m_W)&=& 1,\nonumber \\
C_7 (m_W) &=& \frac{x}{24(x-1)^4}[ 6x  (3x-2)  \ln x  \nonumber \\ 
&-& (x-1)(8x^2 +5x-7)  ],
\end{eqnarray}
with $x=m_t^2/m_W^2$.

Keeping only ${\cal O}_2$ and ${\cal O}_7$ the total $b \rightarrow s
\gamma$ decay rate can be written as 
\begin{equation}
\Gamma = \Gamma_{77} + \Gamma_{22} + \Gamma_{27},
\end{equation}
where $\Gamma_{22}$($\Gamma_{77}$) denotes the decay rate obtained
from the matrix elements squared for ${\cal O}_2$ (${\cal O}_7$) and
$\Gamma_{27}$ is the decay rate from the interference terms between
${\cal O}_2$ and ${\cal O}_7$. Including both virtual correction and
real gluon emission $\Gamma_{77}$ is given by 
\cite{aligreub}
\begin{equation} 
\Gamma_{77} = \Gamma_0 \, C_7(\mu)^2 \, 
\Bigl[ 1 -\frac{2\alpha_s(\mu_R)}{3\pi}(\frac{2\pi^2}{3} 
-\frac{8}{3}) +\cdots \Bigr],  
\label{decay}
\end{equation}
where $\Gamma_0 =|\lambda_t|^2 \alpha G_F^2 m_b^5 / (32\pi^4)$ and
$\lambda_t = V_{tb}V^*_{ts}$. Here we neglect the $s$ quark mass and
the ellipsis denotes higher order contributions in $\alpha_s$. 

To set the generalized BLM scale according to our proposal, it is
necessary to calculate the QCD corrections of order $\alpha_s^2$,
which are proportional to the number $n_f$ of the light quark
flavors. Smith and Voloshin \cite{smvol} have recently shown that the
$n_f$-dependent part of the order $\alpha_s^2$ contribution may be
written in terms of the one-loop corrections evaluated with a
fictitious gluon mass $\lambda$  
\begin{eqnarray}
\delta \Gamma^{(2)} &=& -\frac{b\alpha_s^{(V)} (m_b)}{4\pi} \nonumber
\\ 
&\times&
\int_0^{\infty} \Bigl[\Gamma^{(1)}(\lambda) -\frac{m_b^2}{\lambda^2
+m_b^2}\Gamma^{(1)} (0) \Bigr] \frac{d\lambda^2}{\lambda^2},
\label{smithvoloshin}
\end{eqnarray}
where $b=11-2n_f/3$ and $\Gamma^{(1)}(\lambda)$ is the order
$\alpha_s$ contribution to the decay rate computed with a finite gluon
mass $\lambda$ and $\alpha_s^{(V)}(m_b)$ is the QCD coupling obtained
in the heavy-quark potential scheme. 

In order to obtain $\Gamma_{77}$ we calculate the virtual and
bremsstrahlung corrections with a finite gluon mass coming from ${\cal
O}_7$. The virtual massive gluon 
correction is obtained by calculating diagrams corresponding to the
dressing of the vertex, the initial-state $b$ quark and the
final-state $s$ quark line with a massive gluon. The bremsstrahlung
correction is calculated from the emission of a massive gluon from
either of the external $b, s$ quarks. Sum of the two contributions are
infrared finite. The sum is calculated analytically as 
\begin{eqnarray}
\Gamma_{77}^{(1)} (x)&=& \frac{\alpha_s}{3\pi}\Gamma_0 \Bigl[
\frac{16}{3} - \frac{4}{3}\pi^2 \nonumber \\
&-&16x -\frac{1}{2}x^2 +\frac{2}{3} x^3 
-3x^2 \ln x  \nonumber \\
&+&\frac{x(3x^2 -10x-20)}{\sqrt{x(4-x)}} \tan^{-1}
\frac{(1-x)\sqrt{x(4-x)}}{x(3-x)} \nonumber \\
&+& \frac{x(28+2x-3x^2)}{\sqrt{x(4-x)}} \tan^{-1}
\sqrt{\frac{4-x}{x}} \nonumber \\
&+&8\tan^{-1} \frac{\sqrt{x(4-x)}}{2-x} \tan^{-1}
\sqrt{\frac{4-x}{x}}\Bigr], 
\label{gamma7}
\end{eqnarray}
where $x$ is the dimensionless gluon to $b$ quark mass ratio, $x
\equiv \lambda^2 /m_b^2$ and $\Gamma_0$ is the tree level decay width. 

After carrying out the integration of Eq.~(\ref{gamma7}) over $x$ 
according to Eq.~(\ref{smithvoloshin}), $\Gamma_{77}$ at order
$\alpha_s^2$ is given by 
\begin{eqnarray}
\Gamma_{77} &=& (C_7(\mu))^2\Gamma_0\Bigl[
1-\frac{4\alpha_s^V(m_b)}{3\pi}(\frac{\pi^2}{3}-\frac{4}{3})
\nonumber \\
&\times& \Bigl(
1+\frac{b\alpha_s^V(m_b)}{4\pi} [2.733] \Bigr) \Bigr].
\label{decayone}
\end{eqnarray}

We can apply the same method for the calculation of $\Gamma_{22}$ and
$\Gamma_{27}$. There are no virtual corrections at order
$\alpha_s$. The double differential decay rates at order $\alpha_s$
with a finite gluon mass are given by 
\widetext
\begin{eqnarray}
\frac{d\Gamma_{22}}{dx_q dx_{\gamma}} &=&
\frac{2\alpha_s}{27\pi}\Gamma_0 (C_2(\mu))^2 \times 
\Bigl[ \, 2 \, (1-A)^2 \,
\Bigl( x_g (1-x-x_{\gamma}) +x x_q\Bigr) \nonumber \\
&-& 2(1-A^2)\Bigl\{x_{\gamma}(1-x- x_{\gamma}) + x_g (1+x-x_g) +x_q
(1-x-x_q)  \Bigr\} \nonumber \\
&+&(1+A)^2 \Bigl\{ x_{\gamma} (1+x-x_g)
+\frac{1-x-x_q}{2x}(x_{\gamma}(1-x-x_{\gamma})+x_g
(1+x-x_g))\Bigr\} \nonumber \\
&+&\Bigl\{ -1-2A+3A^2 +\frac{8(1-A)x}{1-x-x_q}-
\frac{4(1-A)^2 x^2}{(1-x-x_q)^2}\Bigr\} \nonumber \\
&\times& \Bigl\{ -x_{\gamma} (1+x-x_g) +\frac{1-x-x_q}{2x} (x_{\gamma} 
(1-x-x_{\gamma}) +x_g (1+x-x_g))\Bigr\}   \Bigl],
\label{gamma22}
\end{eqnarray}
where $x_i = 2E_i/m_b$ ($i=q, \gamma, g$ are $s$ quark, photon, gluon
in the final state) satisfying $x_q+x_{\gamma}+x_g=2$ and
$x=\lambda^2 /m_b^2$. And 
\begin{equation}
A =\frac{x}{1-x-x_q}\ln \frac{1-x_q}{x}.
\end{equation}
Similarly 
\widetext
\begin{eqnarray}
\frac{d\Gamma_{27}}{dx_q dx_{\gamma}}
&=& -\frac{\alpha_s}{9\pi}\Gamma_0 C_2(\mu) C_7(\mu)
\nonumber \\
&\times& \Bigl[ \frac{1}{x-x_g} \Bigl\{ 2(1+x-x_g)(1-x-x_q) +
x_{\gamma} (1-x-x_{\gamma}) + \frac{2x x_{\gamma} (1+x-x_g)}{1-x-x_q}  
\nonumber \\
&-&x_q (1-x-x_q)-x_g (1+x-x_g) -4x(1+x-x_g)
\nonumber \\
&+& A \Bigl( (x_g-4x_{\gamma})(1+x-x_g) +( 4(1+x-x_g)+ x_q) (1-x-x_q) 
-x_{\gamma} (1-x-x_{\gamma})
\nonumber \\
&+&\frac{2 x_g}{x}(1+x-x_g)(1-x-x_q) - 
\frac{2x x_{\gamma}(1+x-x_g)}{1-x-x_q} \Bigr) \Bigr\}
\nonumber \\
&+& \frac{1}{1-x_{\gamma}} \Bigl\{ x_q (1-x-x_q) +x_{\gamma}
(1-x-x_{\gamma}) -x_g(1+x-x_g) + 2x x_{\gamma}\frac{1+x-x_g}{1-x-x_q} 
\nonumber \\
&+& A \Bigl( -2xx_{\gamma} \frac{1+x-x_g}{1-x-x_q}
+\frac{1-x-x_q}{x} ( x_g (1+x-x_g) +x_{\gamma}(1-x-x_{\gamma})-x_q
(1-x-x_q))  \nonumber \\
&-&4x_{\gamma}(1+x-x_g)-x_q (1-x-x_q) -x_{\gamma}(1-x-x_{\gamma})+x_g
(1+r-x_g) \Bigr) \Bigr\} \Bigr].   
\label{gamma27}
\end{eqnarray}
\narrowtext
These are infrared finite as $x\rightarrow 0$ and the phase-space
integration and the integration over the gluon mass yield finite
results. We can get the single differential decay rates by integrating
Eqs.~(\ref{gamma22}) and (\ref{gamma27}) over, say $x_q$, analytically
and the remainingg calculation is done numerically. Using the
Eq.~(\ref{smithvoloshin}) and after the integration, we find
\begin{eqnarray}
\Gamma_{22} &=&  
+ \, C_2^2 \Gamma_0 \frac{4\alpha_s^V(m_b)}{81\pi}
\Bigl[ 1 + \frac{b\alpha_s^V(m_b)}{4\pi}[2.486] + \cdots  \Bigr],  
\label{decaytwo}
\\ 
\Gamma_{27}&=& \! -  C_2 C_7 \Gamma_0
\frac{2\alpha_s^V(m_b)}{27\pi}  \Bigl[  1+\frac{b\alpha_s^V(m_b)}{4\pi}
[6.648] + \cdots \Bigr].
\label{decaythree} 
\end{eqnarray}

Summing Eqs.~(\ref{decayone}), (\ref{decaytwo}), and
(\ref{decaythree}) the total decay rate is given by
\begin{eqnarray}
\Gamma (\mu) &=&\Gamma_{77} (\mu) + \Gamma_{22} (\mu) +
\Gamma_{27}(\mu) \nonumber \\ 
&=& \Gamma_0 (C_7(\mu))^2  \Bigl[  
1 +\frac{\alpha_s}{3\pi}\Bigl(
-7.83 - 0.222 \kappa + 0.148 \kappa^2 \Bigr)
\nonumber \\ 
&\times& \Bigl( 1 + \frac{b\alpha_s}{4\pi} 
 \frac{21.5  + 1.48 \kappa - 0.368 \kappa^2 } 
{7.84 +  0.222 \kappa - 0.148 \kappa^2} 
\Bigl) + \cdots\Bigr],
\label{total}
\end{eqnarray}
where $\kappa = C_2 / C_7$ is the ratio of the Wilson coefficient 
functions. In Eq.~(\ref{total}) the terms proportional to $n_f$ 
can be absorbed 
into the order $\alpha_s$ term according to the BLM prescription. 
Recall that there are two independent scales in the decay rate. 
One is the factorization scale
$\mu$ at which the Wilson coefficient functions and the local
operators are factorized and the other is the renormalization scale
$\mu_R$ at which the theory is renormalized, and the two scales are
independent. 

As we have mentioned, we propose to set the BLM scale only from the
fixed-order perturbation series in $\alpha_s (\mu_R)$. This method
extends the original idea of Brodsky, Lepage and Mackenzie\cite{blm}
closely in the sense that the BLM scale is determined by absorbing 
the vacuum polarizations of light fermions into the coupling constant 
$\alpha_s(\mu_{BLM})$. In this case the coefficients in front of the 
powers of $\alpha_s(\mu_R)$ are functions of the factorization scale 
through the anomalous dimension of the operators. 
Therefore the BLM scale depends on the factorization scale. 
Noting that in the heavy quark potential scheme 
\begin{equation}
\alpha_s^{(V)} (m_b)= \alpha_s^{(V)} (Q) ( 1+ \frac{b}{4\pi}
\alpha_s^{(V)} (Q) \ln \frac{Q^2}{m_b^2} ),   
\end{equation}
the BLM scale from Eq.~(\ref{total}) is given by
\begin{equation}
\mu_{BLM} = m_b\exp \Bigl[-\frac{1}{2}\cdot 
\frac{21.4 +  1.48 \kappa   - 0.368 \kappa^2 }
{7.83 +  0.222 \kappa   -  0.148 \kappa^2 } \Bigr].
\end{equation}
Here $\kappa$ is a function of the factorization scale $\mu$. 
In Fig.~1, the BLM scale $\mu_{BLM}$ is plotted as a function of the 
factorization scale $\mu$. For numerical evaluation we have used $m_t 
= 180$ GeV, $m_W= 80.1$ GeV, $m_b = 5$ GeV and $\Lambda = 225$ MeV for 
$n_f =5$. Our result in Fig.~1 shows that the BLM scale is extremely 
insensitive to the 
factorization scale. As we vary the factorization scale $\mu$ from 
$m_b/2$ to $2m_b$, the BLM scale changes only over $0.315 \, m_b$ to 
$0.334 \, m_b$. 
This is gratifying since the insensitivity of the factorization scale
implies that we can reduce part of the theoretical error of the decay
rate from the choice of the factorization scale.
In our case the renormalization scale is almost constant and it yields
the normalized decay rate which does not vary widely when the
factorization scale changes from $m_b/2$ to $m_b$. 

The total decay rate is then normalized by scaling down the leading
order coefficient functions from $\mu$ to $\mu_{\rm BLM}$ using the
leading logarithmic coefficient function. According to the original
idea of BLM, the normalized rate may be interpreted as the mean total 
decay rate averaged over the gluon virtuality. In order to compare
this normalized total decay rate with other decay rates, we have
plotted three decay rates as a function of the factorization
scale in Fig.~2. The dotted line represents the $\alpha_s^0$-order
result from ${\cal O}_7$ only. The dashed line is the $\alpha_s$ order
result in Eq.~(\ref{total}) with the factorization scale $\mu$ and the
solid line is the $\alpha_s$ result with $\mu = \mu_{BLM}(\mu)$. As
shown in Fig.~2, the result with the BLM scale is 
almost flat in the range while other results vary. Note that in
plotting Fig.~2, we have used the coefficient functions only to the
leading logarithm order. If the complete next-to-leading logarithmic
results for the coefficient functions and the anomalous dimensions are
available in the future, it will be interesting to compare 
the next-to-leading log result of the decay rate with our BLM
result which will also be improved by using the coefficient functions
at next-to-leading orer. This will eventually test
whether our proposal of setting the renormalization scale is
sensible. 

We can also ask questions like how the BLM scales and the decay rate
can be expressed in other schemes. Since there exists a definite
relationship of $\mu_{BLM}$ and physical  
observables among various schemes, our result can be converted without
any ambiguity to other schemes such as the $\overline {MS}$ scheme and
can also be extended to include the running mass effects of the $b$
quark.\cite{lu}   

We thank S. Brodsky, L. Dixon, C. Greub, M. Lautenbacher and M.B. Wise 
for many helpful discussions. This work was supported in part by KOSEF 
Grant 941-0200-02202 (JGC), U.S. NSF-KOSEF Bilateral Grant (SJR), 
KRF Nondirected Grant '94 (SJR), Ministry of Education SNU-BSRI
94-2418 (SJR), KU-BSRI 94-2408 (JGC) and KOSEF SRC Program (JGC, SJR).

\begin{figure}
\caption{The BLM scale $\mu_{BLM}/m_b$ for the total decay rate $\Gamma$ 
as a function of $\mu /m_b$. }
\end{figure}
\begin{figure}
\caption{The total decay rates $\Gamma / \Gamma_0$ as a function of
$\mu/m_b$. } 
\end{figure}
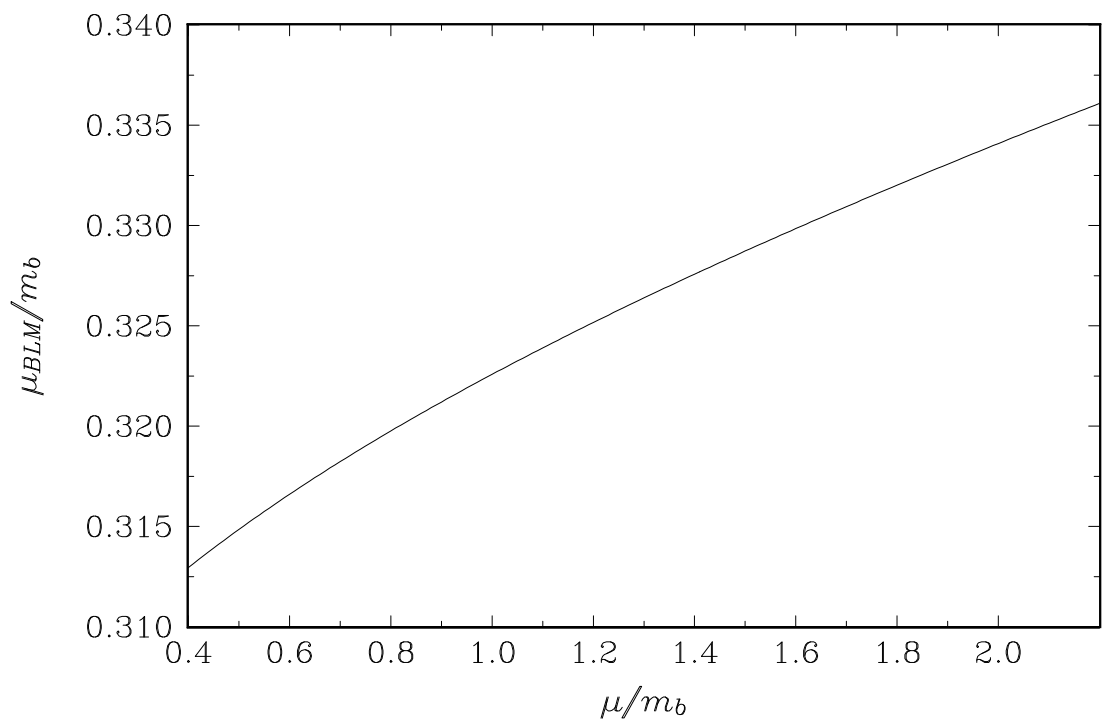
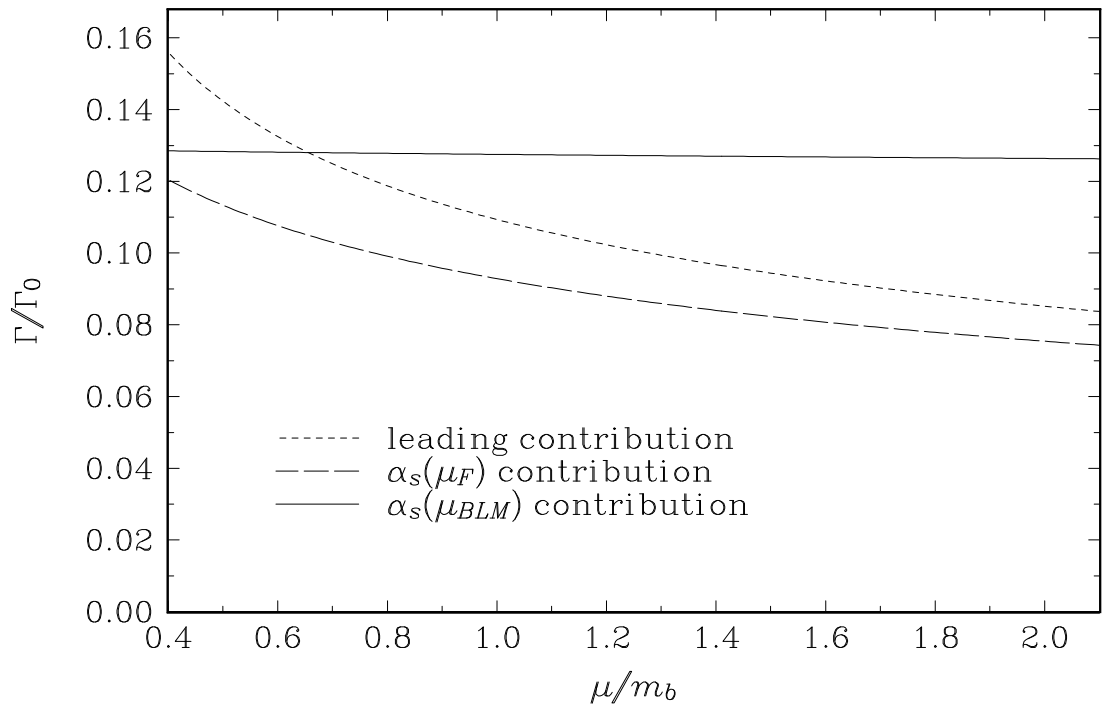
\end{document}